# Superfluid density measurements of Ba(Co$_x$Fe$_{1-x}$)$_2$As$_2$ films near optimal doping

Jie Yong,[1,a)] S. Lee,[2] J. Jiang,[3] C. W. Bark,[2] J. D. Weiss,[3] E. E. Hellstrom,[3] D. C. Larbalestier,[3] C. B. Eom,[2] and T. R. Lemberger[1]

[1]*Department of Physics, The Ohio State University, Columbus, Ohio 43210, USA*

[2]*Department of Materials Science & Engineering, University of Wisconsin-Madison, Madison, Wisconsin, 53706, USA*

[3]*Applied Superconductivity Center, National High Magnetic Field Laboratory, Florida State University, Tallahassee, Florida 32310, USA*

(Received:                                    )

We report the first direct measurements of superfluid density, $n_s(T) \propto \lambda^{-2}(T)$, in films of Fe-pnictide superconductors. The magnetic penetration depth, $\lambda(0)$, in our epitaxial, single-crystal Ba(Co$_x$Fe$_{1-x}$)$_2$As$_2$ films near optimal doping (x=0.08) is 350 nm to 430 nm, comparable to bulk single crystals. The T-dependence of $\lambda^{-2}$ indicates a small s-wave gap, $2\Delta(0)/k_B T_c = 2.2 \pm 0.1$. In detail, $\lambda$ has power-law behavior at low T: $\lambda(T)/\lambda(0) - 1 \approx 0.60*(T/T_c)^{2.5\pm0.1}$. The small gap, together with power-law behavior at low T, suggests strong intraband scattering on the larger-gap Fermi surface and significant interband scattering between large-gap and small-gap Fermi surfaces.

The gap nature of the newly found iron pnictide superconductors has generated great interest in the condensed matter community. Determination of the pairing symmetries and whether nodes exist can give key information on the pairing interaction. $s^{\pm}$ symmetry[1] is proposed by calculations, where the superconducting gaps on hole-like and electron-like pieces of the Fermi surface have opposite signs, but details of the gaps are still uncertain[2,3,4]. There have been a number of measurements on films and crystals of various pnictide superconductors that provide information on the gaps (for a detailed review, see Refs.2 and 3), including several studies on $Ba(Co_xFe_{1-x})_2As_2$ crystals and films discussed here. In this paper, we report the first direct measurement of the superfluid density in an Fe-pnictide films of $Ba(Co_xFe_{1-x})_2As_2$ with nominal doping $x = 0.08$.

Our results on pnictide films are interesting for several reasons. Most generally, in growing films of pnictides, we learn new things about growth and engineering of films of complex materials. In our case, we have found that exceptionally strong vortex-pinning insulating Ba-Fe-O nanopillars under certain growth conditions. [Ref: 5 to 7]. There are a number of important experimental probes that can be applied to films but not on bulk materials, e.g., the experimental technique used here to measure superfluid density. Superfluid density measurements provide an excellent characterization of the quality of superconducting materials, and, when applied to high-quality films, they provide detailed and fundamental information about the superconducting state, as described below.

Table I compares various measurements on $Ba(Co_xFe_{1-x})_2As_2$ (Ba-122) crystals and films.(refs 8-22) Studies 1 through 6 probe the magnetic penetration depth, $\lambda(T)$, either directly or indirectly through the conductivity sum rule, which yields the superfluid density, $n_s \propto \lambda^{-2}$. (As discussed below, $\lambda^{-2}$ is often referred to as the "superfluid density", because the two quantities are proportional. We follow this convention here.) The first five studies also provide the T-dependence of $\lambda^{-2}$, at least at low T. The next two are spectroscopic measurements, ARPES and Point Contact Andreev Reflection (PCAR), which probe the magnitude of the energy gap at low T. The last entry is specific heat.

Quantitatively, $\lambda(0)$ is roughly the same for the narrow range of dopings represented in Table I, ranging from about 325 nm in crystals to about 420 nm in films. However, this range represents a ±25% spread in superfluid densities. This range might be due to different disorder in different samples because microwave measurements have already shown that superfluid density is disorder-dependent in K-doped $BaFe_2As_2$.[23]

The low-T behavior of $\lambda$ is a sensitive probe of the low-energy superconducting density of states. There is agreement among the top five entries in Table I that $\lambda(T)/\lambda(0) - 1 \equiv \Delta\lambda(T)/\lambda(0)$ has power-law behavior at low T, with an exponent $n = 2.3 \pm 0.3$, rather than thermally activated behavior, $exp(-\Delta/k_BT)$. Power-law behavior in $\lambda$ seems to disagree with thermal conductivity measurements[24, 25] in slightly overdoped Ba-122, which show a negligible value for $\kappa_0/T$ as $T/T_c \to 0$, ($\kappa_0$ is the in-plane thermal conductivity), thus providing no evidence of nodes on the Fermi surface. ARPES results[20] also conclude isotropic nodeless gaps.

The experimental situation regarding the energy gap is complicated. As indicated in Table I, some measurements indicate a single small s-wave gap, $2\Delta(0)/k_BT_c \approx 2$, (*i.e.*, studies 1, 5, 6, 7 in Table I) that is only 40% of the BCS minimum value. Others (studies 2, 11, 12 in Table I) indicate the presence of large gaps, $2\Delta(0)/k_BT_c \approx 5\text{-}6$, that are 60% larger than BCS, while still others indicate a mixture of small and large gaps. Thus, it seems that there are two s-wave gaps that differ by a factor of 3 or so, probably residing on different Fermi surface sheets, and that different measurements weight the two gaps differently. This interpretation still leaves open the question as to why different penetration depth studies on nominally similar samples do not see the same gap. We return to this discussion after presenting our data.

We choose to study Ba(Co$_x$Fe$_{1-x}$)$_2$As$_2$[5, 26, 27] because single-crystal epitaxial thin films with superior structural and electromagnetic properties became recently available. Epitaxial Ba(Co$_{0.08}$Fe$_{0.92}$)$_2$As$_2$ thin films are fabricated by pulsed laser deposition onto (001)-oriented (La,Sr)(Al,Ta)O$_3$ (LSAT) substrates precoated with a single-crystal SrTiO$_3$ template layer.[5] Films are deposited in vacuum and are about 100 nm thick. The films have high superconducting and crystalline quality but they do contain a dense array of insulating Ba-Fe-O nanopillars. These nanopillars were characterized by extensive HRTEM[5, 6, 7] which show insulating Ba-Fe-O nanocolumns ~5 *nm* in diameter, spaced about 25 *nm* apart. These nanocolumns are extremely effective vortex pinning defects, but the HRTEM indicates minimal strain in the superconducting matrix, consistent with the very narrow x-ray peaks.[5] Indeed there is no sign of depression of the matrix superconducting properties in any of our extensive studies and we thus expect that the nanopillars will have only a volumetric dilution effect of order 10% on the superfluid density.

A previous study of these films shows that their residual resistance ratio is about 1.5,[5] compared with about 3 for bulk single crystals.[28] Their residual resistivities are about 50% higher than for crystals, suggesting higher disorder. Critical current densities are much larger than for crystals, greater than 1 MA/cm$^2$ at 4.2 K, most likely enhanced by vortex pinning by the insulating nanocolumns.

Superfluid densities are measured by a two-coil mutual inductance apparatus.[29,30] The film is sandwiched between two coils, and the mutual inductance between these two coils is measured at a frequency $\omega/2\pi$ = 50 kHz. The measurement actually determines the sheet conductivity, $Y \equiv (\sigma_1 + i\sigma_2)d$, with $d$ being the superconducting film thickness and $\sigma$ being the conductivity. Given a measured film thickness, $\sigma$ is calculated as: $\sigma = Y/d$. The imaginary part, $\sigma_2$, yields the superfluid density through a low frequency measurement of: $\omega\sigma_2 \equiv n_s e^2/m$, which is proportional to the inverse penetration depth squared: $\lambda^{-2}(T) \equiv \mu_0 \omega \sigma_2(T)$, where $\mu_0$ is the permeability of vacuum.

The dissipative part of the conductivity, $\sigma_1(T)$, has a peak near $T_c$, whose width provides an upper limit on the inhomogeneity of $T_c$. Measurements are taken continuously as the sample warms up so as to yield the hard-to-measure absolute value of $\lambda$ and its T-dependence, which sheds light on the superconducting energy gap.

Five films on either bare $SrTiO_3$ substrates or on $SrTiO_3$ template layers on LSAT substrates were measured as detailed in Table II. We will not discuss data on films that were deposited onto bare LSAT substrates because those films were of significantly lower quality due to their poor epitaxy which gave much lower superfluid densities and critical current densities.[5] Figure 1 shows $\lambda^{-2}(T)$ for two typical $Ba(Co_{0.08}Fe_{0.92})_2As_2$ films, films A and B of Table II. $\lambda(0)$ ranges from 350 nm (A) to 430 nm (B to E) in our films (Table II). These values are slightly larger than values in crystals possibly due to higher disorder.

Narrow peaks of $\sigma_1$ indicate the good homogeneity of $T_c$ in our films. As seen in Fig. 1, the overall T-dependence of $\lambda^{-2}$ is fitted well to 1% or so by BCS theory in the dirty limit[31], except for a small tail extending above the "$T_c$" obtained from the fit. We emphasize that the fit includes just one s-wave gap that is much smaller, $2\Delta/k_B T_c \sim 2.2$, than the value, 3.53, expected for weak-coupling s-wave BCS superconductors. The other two fit parameters, $\lambda^{-2}(0)$ and $T_c$, have no influence on the T-dependence of the fit. Thus, $\lambda^{-2}(T)$ indicates that essentially all of the superfluid comes from Fermi surface sheets with a small s-wave gap. A similar argument has been made to explain the anomalous magnetic-field dependence of $\kappa_0/T$ in overdoped Ba-122.[25] A small gap also accounts for most of the spectral "missing area" in optical measurements.[19]

There are two minor discrepancies in the fit. (1) The first few percent drop at low T is better fitted by a power law: $\Delta\lambda(T)/\lambda(0) = 0.60*(T/T_c)^{2.5\pm0.1}$ than by thermally activated behavior (see the insets in Fig. 1). This power-law deviation from activated behavior in the data is less than 1%, but is experimentally clear and manifests in all of our samples and is also found in crystals.[11, 12] In fact, films and crystals agree remarkably well in that both the exponent and the normalized coefficient are close, as shown in Table I. (2) There is a small foot that extends above "$T_c$". This could be due to slight sample inhomogeneity or it could be an indication of a small amount of superfluid associated with a large gap. A similar tail has also been observed in TDR[12] and microwave[14] measurements. This feature is the only possible indication of the presence of a large gap in our films.

The superfluid density of films B to E are very close, whereas that of film A is about 40% higher (see Table II). The power law exponent for Film E is n ≈ 2.26, slightly smaller than those of the other four films. This might mean that film E is slightly underdoped, given its relatively lower $T_c$. This slight drop in exponent is consistent with TDR[12] and local MFM[8] measurements on 5% Co-doped crystals. Also, the foot near $T_c$ is absent for film E, which might mean that the gap nature changed due to a lower doping, for example, significant in-plane anisotropy has been observed for underdoped Ba-122 due to a tetragonal to orthorhombic structure change.[32]

Now let us discuss the physics of the results. First, the small gap and power-law behavior at low T indicate a non-BCS density of states that peaks at an energy near $k_BT_c$, but extends to lower energies to account for the low-T power-law behavior. Elastic interband scattering may be the explanation. Density functional theory predicts that iron pnictides likely have $s^{\pm}$ symmetry, where the hole-like Fermi surfaces centered on the $\Gamma$ point and the electron-like Fermi surfaces centered on the M point have s-wave gaps of opposite sign.[1] In this case, theory[33] predicts that scattering between hole-like and electron-like Fermi surfaces smears out the square root singularity in the BCS density of states analogously to the effect of magnetic impurities on conventional superconductors,[34] and thereby changes the low-T behavior of $\lambda^{-2}(T)$ from thermally activated to power law, with an exponent between 2 and 3 depending on the strength of interband scattering. The common observation of a low-T exponent near 2.5 in films and crystals suggests both the importance of interband impurity scattering, and that the exponent is likely to be somewhat insensitive to the elastic scattering rate, to the extent that films and crystals have different degrees of disorder.

In this context, it is worth noting that some oxygen-containing iron-based superconductors like $PrFeAsO_{1-y}$ single crystals[35] show an activated T-dependence of the penetration depth. This is consistent with their higher $T_c$'s, assuming that the $T_c$ of Ba-122 compounds is lowered below that of $PrFeAsO_{1-y}$ solely by stronger interband scattering. Such arguments assume that all iron-based superconductors have the same pairing symmetry due to their structural similarities. However, we cannot rule out other pairing symmetries, for example, that the gap on the Fermi surface near the M point has accidental nodes or has d-wave symmetry, because power-law behavior might also be due to nodes on the Fermi surface.

Second, it is puzzling that different measurements of $\lambda(T)$ find widely different energy gaps. For example, $\mu SR$[10] and MFM[8] measurements indicate that some superfluid comes from Fermi surface regions with a small s-wave gap while most of it comes from regions with a relatively large s-wave gap, $2\Delta/k_BT_c \approx 5$ or 3.8, respectively. On the other hand, microwave measurements[14] find that all of the superfluid is associated with a single small gap, $2\Delta/k_BT_c \approx 2$. As for films, both our measurements and the low-T optical conductivity measurements[15-19] listed in Table I find the same small gap as do the microwave measurements. One explanation of this missing "large-gap" superfluid density is that the intraband elastic scattering on the large-gap Fermi surface in some samples is strong enough to completely suppress the large-gap superfluid, so that only the small gap superfluid is observed. However, such an explanation has a problem that the superfluid density in such disordered samples should be several times smaller than in the cleaner samples, and there is no evidence for such in the data.

To summarize, we measured superfluid densities of near optimally doped $Ba(Co_{0.08}Fe_{0.92})_2As_2$ films. We find that: (1) $\lambda(0)$ is about 420 nm, which implies a superfluid density about two-thirds of that in crystals. Some of the difference may arise because the insulating Ba-Fe-O nanorods dilute the superfluid density by 10% or so. (2) Most of the superfluid density comes from a Fermi surface with a small s-wave gap, $2\Delta/k_BT_c = 2.2\pm0.1$. (3) The low T behavior of $\lambda(T)$ exhibits power-law behavior, $\Delta\lambda(T)/\lambda(0) = 0.6*(T/T_c)^{2.5\pm0.1}$, in detailed agreement with measurements of $\lambda(T)$ in other films and in crystals. This power-law behavior, together with the observation of only a small, sub-BCS gap, indicates a non-BCS density of states that we believe is a consequence of an $s^{\pm}$ gap symmetry and strong interband scattering. (4) The only possible evidence for a large-gap superfluid in our measurements is the slight tail in $\lambda^{-2}$ near $T_c$.

We find no evidence that the insulating nanopillars in our films, which are such effective pinners of c-axis vortices, have a significant effect on the magnitude or T-dependence of the superfluid density, other than by producing a proportional volumetric reduction of order 10%.

The ability to make accurate measurements of $\lambda(T)$ on films enables measurements of superfluid density to become a key parameter in characterizing the quality of Fe-pnictide films. As multiple groups are now growing good quality films in different ways we may expect that more measurements will become available to clarify some of the residual uncertainties noted above.

Acknowledgements: This work was supported by U.S. Department of Energy, Office of Basic Energy Sciences, under Grants FG02-08ER46533 at OSU and DE-FG02-06ER46327 at UW-Madison. The work at the NHMFL was supported under NSF Cooperative Agreement No. DMR-0084173, by the State of Florida, and by AFOSR under Grant No. FA9550-06-1-0474.

**References:**

[1] I. I. Mazin, D. J. Singh, M. D. Johannes, and M. H. Du, Phys. Rev. Lett, 101, 057003 (2008).

[2] J. Paglione and R.L.Greene, Nature Physics, 6, 645 (2010).

[3] D. C. Johnston, arXiv:1005.4392v2.

[4] I. I. Mazin, T. P. Devereaux, J. G. Analytis, Jiun-Haw Chu, I. R. Fisher, B. Muschler, and R. Hackl, Phys. Rev. B 82, 180502 (R) (2010).

[5] S. Lee, J. Jiang, Y. Zhang, C. W. Bark, J. D. Weiss, C. Tarantini, C. T. Nelson, H. W. Jang, C. M. Folkman, S. H. Baek, A. Polyanskii, D. Abraimov, A. Yamamoto, J. W. Park, X. Q. Pan, E. E. Hellstrom, D. C. Larbalestier, and C. B. Eom, Nature Mater. **9**, 397 (2010).

[6] C. Tarantini, S. Lee, Y. Zhang, J. Jiang, C. W. Bark, J. D. Weiss, A. Polyanskii, C. T. Nelson, H. W. Jang, C. M. Folkman, S. H. Baek, X. Q. Pan, A. Gurevich, E. E. Hellstrom, C. B. Eom, and D. C. Larbalestier, Appl. Phys. Lett. **97**, 022506 (2010).

[7] Y. Zhang, C.T. Nelson, S.Lee, J. Jiang, C.W.Bark, J. D. Weiss, C. Tarantini, C.M. Folkman, S.H.Baek, E.E.Hellstrom, D. Larbalestier, C.B. Eom, X. Pan, accepted for publication in APL

[8] L. Luan, O. M. Auslaender, T. M. Lippman, C.W. Hicks, B. Kalisky, J.H. Chu, J. G. Analytis, I. R. Fisher, J. R. Kirtley, and K. A. Moler, Phys. Rev. B 81, 100501(R) (2010).

[9] L.Luan, T.M. Lippman, C.W. Hicks, J.A.Bert, O.M. Auslaender, J. H. Chu, J. G. Analytis, I. R. Fisher and K.A. Moler, arXiv:1012.3436v1

[10] T. J. Williams, A. A. Aczel, E. Baggio-Saitovitch, S. L. Bud'ko, P. C. Canfield, J. P. Carlo, T. Goko, J. Munevar, N. Ni, Y. J. Uemura, W. Yu, and G. M. Luke, Phys. Rev. B, 80, 094501 (2009).

[11] R. T. Gordon, N. Ni, C. Martin, M. A. Tanatar, M.D. Vannette, H. Kim, G. D. Samolyuk, J. Schmalian, S. Nandi, A. Kreyssig, A. I. Goldman, J. Q. Yan, S. L. Bud'ko, P. C. Canfield, and R. Prozorov Phys. Rev. Lett. **102**, 127004 (2009)

[12] R. T. Gordon, C. Martin, H. Kim, N. Ni, M. A. Tanatar, J. Schmalian, I. I. Mazin, S. L. Bud'ko, P. C. Canfield, and R. Prozorov, Phys. Rev. B **79**, 100506(R) (2009)

[13] R. T. Gordon, H. Kim, N. Salovich, R. W. Giannetta, R. M. Fernandes, V. G. Kogan, T. Prozorov, S. L. Bud'ko, P. C. Canfield, M. A. Tanatar, and R. Prozorov Phys. Rev. B 82, 054507 (2010)

[14] J. S. Bobowski, J. C. Baglo, James Day, P. Dosanjh, Rinat Ofer, B. J. Ramshaw, Ruixing Liang, D. A. Bonn, and W. N. Hardy, Phys. Rev. B 82, 094520 (2010).

[15] B. Gorshunov, D. Wu, A. A. Voronkov, P. Kallina, K. Iida, S. Haindl, F. Kurth, L. Schultz, B. Holzapfel and M. Dressel, Phys. Rev. B 81, 060509 (R) (2010).

[16] A. Perucchi, L. Baldassarre, C. Marini, S. Lupi, J. Jiang, J. D. Weiss, E.E. Hellstrom, S. Lee, C.W. Bark, C. B. Eom, M. Putti, I. Pallecchi, and P. Dore, Eur. Phys. J. B **77**, 25 (2010)

[17] K. W. Kim, M. Rössle, A. Dubroka, V. K. Malik, T. Wolf, and C. Bernhard, Phys. Rev. B 81, 214508 (2010).

[18] E. Van Heumen, Y. Huang, S. de Jong, A.B.Kuzmenko, M. S. Golden and D. van der Marel, Euro. Phys. Lett, 90, 37005 (2010).

[19] J. J. Tu, J. Li, W. Liu, A. Punnoose, Y. Gong, Y. H. Ren, L. J. Li, G. H. Cao, Z. A. Xu and C. C. Homes, Phys. Rev. B 82, 174509 (2010).

[20] K. Terashima, Y. Sekiba, J. H. Bowen, K. Nakayama, T. Kawaharab, T. Sato, P. Richard, Y.-M. Xu, L. J. Li, G. H. Cao, Z.-A. Xu, H. Ding, and T. Takahashi, Proc. Natl. Acad. Sci. 106, 7330 (2009).

[21] P. Samuely, Z. Pribulova, P. Szabo, G. Prista, S.L. Bud'ko, P.C. Canfield, Physica C. 469, 507 (2009).

[22] F. Hardy, T. Wolf, R. A. Fisher, R. Eder, P. Schweiss, P. Adelmann, H. v. Löhneysen, and C. Meingast. Phys. Rev B 81, 060501 (R) (2010)

[23] K. Hashimoto, T. Shibauchi, S. Kasahara, K. Ikada, S. Tonegawa, T. Kato, R. Okazaki, C. J. van der Beek, M. Konczykowski, H. Takeya, K. Hirata, T. Terashima, and Y. Matsuda, Phys. Rev. Lett. 102, 207001 (2009)

[24] M. A. Tanatar, J.-Ph. Reid, H. Shakeripour, X. G. Luo, N. Doiron-Leyraud, N. Ni, S. L. Bud'ko, P. C. Canfield, R. Prozorov, and Louis Taillefer Phys. Rev. Lett, 104, 067002 (2010).

[25] J. K. Dong, S. Y. Zhou, T. Y. Guan, X. Qiu, C. Zhang, P. Cheng, L. Fang, H. H. Wen, and S. Y. Li, Phys. Rev. B 81, 094520 (2010).

[26] T. Thersleff, K. Iida, S. Haindl, M. Kidszun, D. Pohl, A. Hartmann, F. Kurth, J. Hänisch, R. Hühne, B. Rellinghaus, L. Schultz, and B. Holzapfel, Appl. Phys. Lett. 97, 022506 (2010)

[27] T. Katase, H.Hiramatsu, H.Yanagi, T.Kamiya, M. Hirano, H. Hosono Solid State Comm.149, 2121 (2009).

[28] A. S. Sefat, Rongying Jin, Michael A. McGuire, Brian C. Sales, David J. Singh, and David Mandrus, Phys. Rev. Lett. 101, 117004 (2008).

[29] S. J. Turneaure, E. R. Ulm, and T. R. Lemberger, J. Appl. Phys. 79, 4221 (1996).11

[30]S. J. Turneaure, A. A. Pesetski and T. R. Lemberger, J. Appl. Phys. 83, 4334 (1998).

[31]T.R.Lemberger, I. Hetel, J. W. Knepper, and F. Y. Yang, Phys. REV. B 76, 094515 (2007).

[32]J. H. Chu, J. G. Analytis, K. De Greve, P. L. McMahon, Z. Islam, Y. Yamamoto and I. R. Fisher, Science, 329, 824 (2010)

[33]Yunkyu Bang Euro. Phys. Lett, 86, 47001 (2009).

[34] A. B. Vorontsov, M. G. Vavilov, and A. V. Chubukov Phys. Rev. B 79, 140507(R) (2009).

[35] K. Hashimoto, T. Shibauchi, T. Kato, K. Ikada, R. Okazaki, H. Shishido, M. Ishikado, H. Kito, A. Iyo, H. Eisaki, S. Shamoto, and Y. Matsuda Phys. Rev. Lett. 102, 017002 (2009).

Table I: Results of various measurements on Ba(Co$_x$Fe$_{1-x}$)$_2$As$_2$. In the 2Δ(0)/k$_B$T$_c$ column, percentages in parentheses indicate the contribution of that gap to the measurement. "main" means that gap dominates that measurement.

| Study | Measurement | Co doping $x$; Sample | λ(0) (nm) | $\frac{\lambda(T)}{\lambda(0)} - 1$ [at low T] | 2Δ(0)/k$_B$T$_c$ |
|---|---|---|---|---|---|
| **1** | **Two-coil, λ (present work)** | **8% films** | **350 - 430** | **0.6*(T/T$_c$)$^{2.55}$** | **2.2 ± 0.1** |
| 2 | MFM,[8,9] λ | 5% crystals | 325±50 | 0.26*(T/T$_c$)$^{2.2}$ | 1.4 and 5.0 (90%) |
| 3 | μSR,[10] λ | 7.4% crystals | 300 | ∝ (T/T$_c$)$^{1.9\pm0.3}$ | 1.6 and 3.8 (70%) |
| 4 | TDR,[11-13] λ | 7% crystals | 270±100 | 0.5*(T/T$_c$)$^{2.4}$ | N.A. |
| 5 | μwave,[14] λ | 5% crystals | N.A. | ∝ (T/T$_c$)$^{2.5}$ | 2.0 |
| 6 | Infrared (IR) Conductivity, σ$_1$(ω,T) | 10% films[15] | 360±50 | N.A. | 2.1±0.1 |
| 7 | | 8% films[16] | N.A. | N.A. | 2 (80%) and 7 |
| 8 | | 6.5% crystals[17] | ~270 | N.A. | 3.1 (50%), 4.7 (>40%), 9 (<10%) |
| 9 | | 7% crystals[18] | 340±30 | N.A. | 3.1 (main) and 7 |
| 10 | | 7.5% crystals[19] | 300±30 | N.A. | 2.9 (main) and 7.3 |
| 11 | ARPES[20] | 7.5% crystals | N.A. | "isotropic gaps" | 5.0 and 6.0 |
| 12 | Andreev Reflection[21] | 7% crystals | N.A. | N.A. | 5.8 ± 0.5 |
| 13 | Spec. Heat[22] | 7.5% crystals | N.A. | N.A. | 1.9 and 4.4 (66%) |

Table II: Penetration depth and other fitting parameters for films A to E. They show $\lambda(0)$ ranges from 350 to 430 nm, with most values near the latter. A small s-wave gap, $2\Delta(0)/k_B T_c = 2.2 \pm 0.1$, contributes most of the superfluid. "n" is the power-law exponent that describes the first few percent decrease in $\lambda^{-2}$ at low T. See text.

| Ba-122 Film | Template/substrate | $\lambda(0)$ (nm) | $2\Delta(0)/k_B T_c$ from BCS fit | Low-T exponent, n | Transition Temperatures | |
|---|---|---|---|---|---|---|
| | | | | | $T_c$ from BCS fit to $\lambda^{-2}$ | $T_c$ from resistivity |
| A | 100uc SrTiO$_3$ /LSAT | 350 | 2.23 | 2.54 | 16.7 K | 17.7 K |
| B | | 430 | 2.16 | 2.55 | 16.2 K | 17.3 K |
| C | | 435 | 2.32 | 2.53 | 16.9 K | 19.5 K |
| D | SrTiO$_3$ | 425 | 2.26 | 2.57 | 17.5 K | 18.8 K |
| E | | 425 | 2.06 | 2.26 | 15.3 K | 15.3 K |

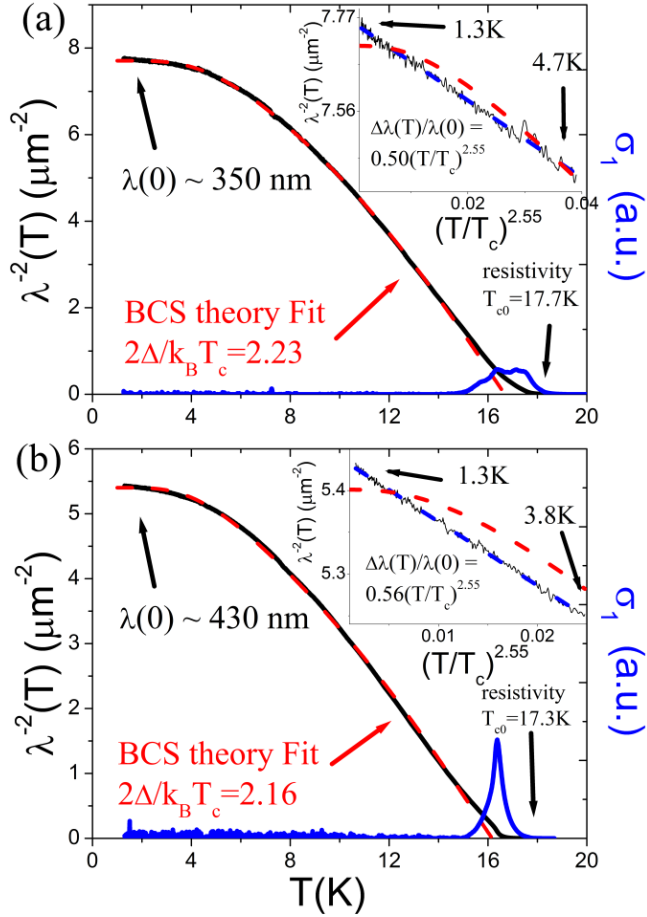

Figure 1. (Color online) (a) and (b): $\lambda^{-2}(T)$ (smooth black curves) and $\sigma_1$ (blue peaks) of $Ba(Co_{0.08}Fe_{0.92})_2As_2$ films A and B, respectively. Dirty-limit BCS theory, with $2\Delta/k_BT_c$ = 2.30 and 2.15, respectively (red dashed curves), fits the data well, except for a <1% discrepancy at low T (insets) and a foot that extends above the fitted $T_c$. The insets show that the low-T behavior is power law, $\Delta\lambda^{-2}(T) \approx AT^b$, where b ≈ 2.55 (blue dashed lines) rather than activated, $exp(-\Delta/k_BT)$ (red dashed curves).